\documentclass[%
 reprint,
 amsmath,amssymb,
 aps,
]{revtex4-2}

\usepackage{graphicx}
\usepackage{dcolumn}
\usepackage{bm}
\usepackage{hyperref}

\newtheorem{theorem}{Theorem}

\newtheorem{definition}{Definition}
\newtheorem{lemma}{Lemma}

\newtheorem{conjecture}{Conjecture}

\usepackage[normalem]{ulem}
\usepackage{color}

\newcommand{\hil}{\mathcal{H}}
\newcommand{\E}{\mathcal{E}}
\newcommand{\sysa}{{\mathcal{S}_A}}
\newcommand{\sysb}{{\mathcal{S}_B}}
\newcommand{\refa}{{\mathcal{R}_A}}
\newcommand{\refb}{{\mathcal{R}_B}}
\newcommand{\refs}{{\rho_\mathcal{R}}}

\newcommand{\schn}{\mathrm{N}_\mathrm{Sch}}
\newcommand{\iidd}{{i=1,\ldots,d^2}}

\newcommand{\op}[1][]{{\left[ {#1} \right]}}
\newcommand{\braket}[2]{\langle #1 | #2 \rangle}
\newcommand{\ketbra}[1][]{| #1 \rangle \langle #1 |}
\newcommand{\ket}[1][]{| #1 \rangle}
\newcommand{\bra}[1][]{\langle #1 |}
\newcommand{\dket}[1][]{| #1 \rangle \rangle}
\newcommand{\dbra}[1][]{\langle \langle #1 |}

\newcommand{\linop}[1][]{\mathbf{L}\left( #1 \right)}
\newcommand{\vspan}[1]{\mathrm{span}\left( #1 \right)}
\newcommand{\range}[1]{\mathrm{range}\left( #1 \right)}

\newcommand{\trace}{\mathrm{Tr}}
\newcommand{\id}{\mathbb{I}}
\newcommand{\qed}{\hfill\blacksquare}

\begin{document}

\preprint{APS/123-QED}

\title{Localization of joint quantum measurements on \ensuremath{\mathbb{C}^d \otimes \mathbb{C}^d} by entangled resources with Schmidt number at most \ensuremath{d}} 

\author{Seiseki Akibue}
 \altaffiliation{These authors contributed equally to this work}
\affiliation{%
Communication Science Laboratories, NTT, Inc.,\\
NTT Research Center for Theoretical Quantum Information,\\
NTT Institute for Fundamental Mathematics,\\
3--1 Morinosato Wakamiya, Atsugi, Kanagawa 243-0198, Japan}

\author{Jisho Miyazaki}%
 \altaffiliation{These authors contributed equally to this work}
\affiliation{%
Graduate School of Science, The University of Tokyo, 7--3--1, Hongo, Bunkyo-ku, Tokyo 113-0033, Japan\\
Ritsumeikan University BKC Research Organization of Social Sciences,
1--1--1, Noji-Higashi, Kusatsu, Shiga 525-8577, Japan
}%

\date{\today}

\begin{abstract}
Localizable measurements are joint quantum measurements that can be implemented using only non-adaptive local operations and shared entanglement. We provide a protocol-independent characterization of localizable projection-valued measures (PVMs) by exploiting algebraic structures that any such measurement must satisfy.
We first show that a rank-1 PVM on $\mathbb{C}^d\otimes\mathbb{C}^d$ containing an element with the maximal Schmidt rank can be localized using entanglement of a Schmidt number at most $d$ if and only if it forms a maximally entangled basis corresponding to a nice unitary error basis. This reveals strong limitations imposed by non-adaptive local operations, in contrast to the adaptive setting where any joint measurement is implementable. We then completely characterize two-qubit rank-1 PVMs that can be localized with two-qubit entanglement, resolving a conjecture of Gisin and Del Santo, and finally extend our characterization to ideal two-qudit measurements, strengthening earlier results.
\end{abstract}

\maketitle
\section{Introduction}


Localizable quantum measurements, adopted in Gisin and Del Santo \cite{Gisin_2024towardsmeasurement}, are a class of multipartite measurements that can be implemented by local operations without any inter-party communication.
Thus, the local operations considered here can be performed instantaneously, i.e., as spacelike-separated events.
The pre-shared entanglement among the parties allows for a non-trivial class of measurements, beyond mere independent local measurements, to be localized.

Localizable measurements started to be researched early in the relativistic context \cite{Aharonov_1980,Aharonov_1981,Aharonov_1986,Popescu_Vaidman_1994,Beckman_2002prd}.
They are central to the long-standing question of what observables on spacelike separated regions should be in relativistic quantum field theory \cite{Gisin_2024towardsmeasurement}, a discussion tracing back to Sorkin's impossible measurement \cite{Sorkin_1993} (see \cite{Fraser_2023} for the history).
This foundational motivation has driven efforts to identify localizable measurements and to build concrete localization protocols.

The localizability of a measurement highly depends on whether it is an ideal measurement or a Positive-Operator-Valued-Measure (POVM), i.e., whether the projected post-measurement states are required or not. Localizable ideal measurements are known to be quite restricted, with complete identification achieved only for two-qubit systems \cite{Popescu_Vaidman_1994,Beckman_2001}. In sharp contrast, if the post-measurement state is disregarded, Vaidman demonstrated a protocol able to localize any POVM by sharing unlimited entanglement resources \cite{Vaidman_2003}. 

This has revealed an unexpected capability of the local operations assisted by entanglement, considering that quantum teleportation, which is a standard subroutine for implementing joint operations by consuming entanglement, relies on classical communication and adaptive local operations and therefore cannot be realized within the instantaneous setting.
This has stimulated subsequent research in quantum information theory, driven by the ubiquity of joint POVMs in quantum information processing and by the fact that local operations are easier to implement in practice than adaptive protocols, which often require quantum memory to store quantum states while awaiting classical messages from other parties. 
From a practical perspective, refinement of Vaidman’s protocols and reduction of the entanglement consumption required for localization have been extensively studied \cite{Groisman_2003optics,Clark_2010,Beigi_2011}.
Despite these efforts, it remains unknown whether an arbitrary POVM can be implemented using only a finite amount of entanglement.


Motivated by this problem, Pauwels \emph{et al.} have systematically investigated localizable POVMs that can be implemented by a given amount of entanglement \cite{pauwels_2025classification}. While their work has successfully characterized the set of POVMs that are localizable under specific classes of local operations, a complete characterization of POVMs that cannot be localized with $N$ ebits---independent of the chosen protocol---has remained elusive, even in the simplest case $N=1$ (see \emph{Note added} in \cite{Pauwels2024_arXiv}).

In this work, we obtain a protocol-independent characterization of localizable measurements by exploiting the algebraic structure that any localizable projection-valued measure (PVM) must satisfy.
As our first result, we show that a rank-1 PVM on $\mathbb{C}^d\otimes\mathbb{C}^d$ containing at least one element with the maximal Schmidt rank can be localized by an entangled state with a Schmidt number at most $d$ if and only if it forms a maximally entangled basis corresponding to a nice unitary error basis (Theorem \ref{thm:lcalizable=nice}), which has been extensively studied in the context of quantum error correction~\cite{knill1996non,Klappenecker,Klappenecker_2005}.
This result not only completely characterize a wide range of localizable PVMs with a bounded amount of entanglement but also reveals implementable joint measurements are strongly restricted by the non-adaptivity of local operations. Indeed, in contrast, any joint measurement can be implemented using a maximally entangled state of Schmidt rank $d$ when one-way classical communication and adaptive local operations are allowed.
As our second result, we completely characterize two-qubit rank-1 PVMs that can be localized using two-qubit entanglement (Theorem \ref{thm:2Qlocalizable}). This strengthens Theorem 1 of Pauwels \emph{et al.}~\cite{pauwels_2025classification} and fully resolves the conjecture posed by Gisin and Del Santo \cite{Gisin_2024towardsmeasurement}.
As our third result, we show that an ideal measurement on a two-qudit basis that has at least one element with the maximal Schmidt rank can be localized by an entangled state (without the assumption on its Schmidt number) if and only if it forms a maximally entangled basis corresponding to a nice unitary error basis (Theorem \ref{thm:Beckman}), thereby strengthening Theorem~6 of Beckman \emph{et al.}~\cite{Beckman_2001}.

\paragraph*{Assumptions and scope.} 
To clarify the applicability of our results, we summarize theorems and lemmas together with their underlying assumptions on the localization scenario in Table~\ref{tab:assumptions}.
Although our primary focus is on cases where the Schmidt number of resource states is bounded relative to the system sizes, several results hold independently of this constraint.
Theorem \ref{thm:equivalence_to_nice} is excluded from Table \ref{tab:assumptions} because it concerns unitary error bases and is formulated independently of the localization problem.
\begin{table*}[htbp]
\caption{\label{tab:assumptions} Summary of results and their underlying assumptions. The condition ``maximal Schmidt rank'' implies that at least one measurement operator possesses the maximal Schmidt rank. $\schn$ denotes the Schmidt number.}
    \begin{ruledtabular}
        \begin{tabular}{lcccc}
            Theorem/Lemma &
            System dimensions &
            Target measurement &
            Resource state $\refs$ \\
            \colrule
            Lemma \ref{lem:rank-1_non-redundant} & arbitrary & POVM & arbitrary \\
            Lemma \ref{lem:closeness_localizable_POVM} & arbitrary & POVM & arbitrary \\
            Lemma \ref{lem:pure} & arbitrary & rank-$1$ PVM & arbitrary \\
            Lemma \ref{lem:rank-1_localizability} & arbitrary & rank-$1$ PVM & arbitrary \\
            Lemma \ref{lem:SchmidtNbound_implies_equalsize} & arbitrary & rank-$1$ PVM & $\schn(\refs) \leq d$ \\
            Theorem \ref{thm:lcalizable=nice} & $\dim \sysa = \dim \sysb = d$ & rank-$1$ PVM, maximal Schmidt rank & $\schn(\refs) \leq d$ \\
            Theorem \ref{thm:2Qlocalizable} & $\dim \sysa = \dim \sysb = 2$ & rank-$1$ PVM & $\schn(\refs) \leq 2$ \\
            Theorem \ref{thm:Beckman} & $\dim \sysa = \dim \sysb = d$ & rank-$1$ ideal projective, maximal Schmidt rank & arbitrary
        \end{tabular}
    \end{ruledtabular}
\end{table*}

\section{Preliminaries and notation}
We denote $A\propto B$ if there exists $\alpha\in\mathbb{C}$, $A=\alpha B$ for two linear operators $A$ and $B$. Note that the definition is not symmetric under the interchange of $A$ and $B$ when $A$ or $B$ is equal to zero. Indeed, $A \propto 0$ holds if and only if $A=0$. In contrast, $0 \propto A$ holds for any $A$.

\subsection{Localizable POVMs}
\begin{figure}[b]
\includegraphics[width=.4\textwidth]{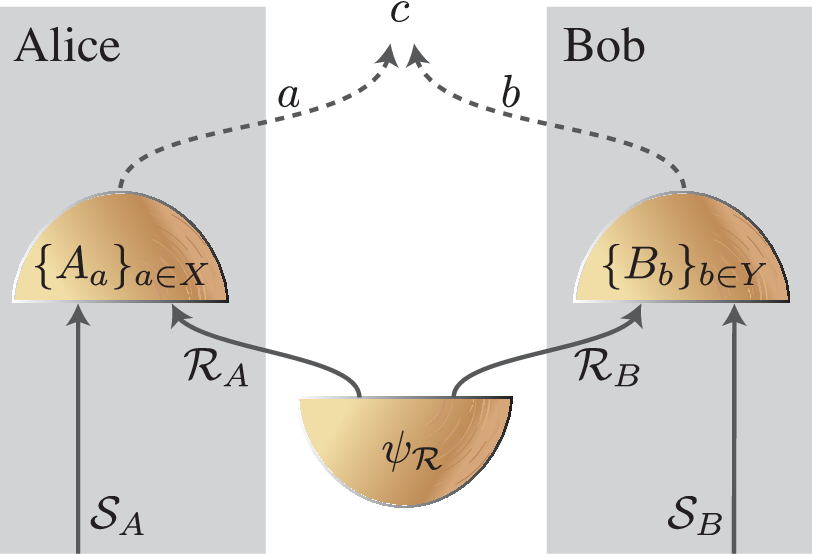}
\caption{\label{fig:localizability_def} The localization scheme for a POVM measurement on $\sysa \otimes \sysb$, depicted by a circuit with transformations applied from bottom to top. The solid and dashed lines represent quantum and classical registers, respectively.}
\end{figure}

The formal definition of localizable joint POVMs includes four subsystems $\sysa,~\sysb,~\refa$ and $\refb$, where the target POVM and the resource state belong to $\sysa \otimes \sysb$ and $\refa \otimes \refb$, respectively.
Alice and Bob, respectively, hold systems $\sysa \otimes \refa$ and $\sysb \otimes \refb$, on which they can perform any local operation.
\begin{definition}[localization]\label{def:localization}
    A POVM $\{ M_c \}_{c \in Z}$ on a joint system $\sysa \otimes \sysb$ with finite $Z$ is defined to be \emph{localizable} by state $\refs$ on $\refa \otimes \refb$ if there exist POVM $\{ A_a \}_{a \in X}$ on $\sysa \otimes \refa$, POVM $\{ B_b \}_{b \in Y}$  on $\sysb \otimes \refb$ and a conditional probability $p(Z|XY)$ such that
    \begin{eqnarray}
        \label{eq:localization} M_c = \sum_{a,b} p(c|a,b) \trace_{\refa,\refb} \left[ (A_a \otimes B_b )~ (\id_{\sysa \otimes \sysb} \otimes \refs ) \right],
    \end{eqnarray}
    holds for all $c \in Z$.
    The tuple $\left( \refs,~ \{ A_a \}_{a \in X},~ \{ B_b \}_{b \in Y},~ p(Z|XY) \right)$ is referred to as a \emph{localization} of $\{ M_c \}_{c \in Z}$.
\end{definition}
The defining scheme of localization is illustrated in Fig.~\ref{fig:localizability_def}.
Note that $p(Z|XY)$ represents a classical post-processing of measurement outcomes.
When the resource state $\refs$ is an $n$-ebit Bell state, the above defined localizability reduces to the $n$-ebit localizability studied by Pauwels \emph{et al.} \cite{pauwels_2025classification}.

As noted in \cite{pauwels_2025classification}, this definition arises from the requirement that the localization reproduce the Born statistics of the target POVM for any input state on $\sysa \otimes \sysb$. Condition \eqref{eq:localization} is equivalent to the requirement that
\begin{eqnarray*}
    \trace[M_c \rho_\mathcal{S}] = \sum_{a,b} p(c|a,b) \trace_{\refa,\refb} \left[ (A_a \otimes B_b )~ (\rho_\mathcal{S} \otimes \refs ) \right],
\end{eqnarray*}
holds for an arbitrary input state $\rho_\mathcal{S}$ on $\sysa \otimes \sysb$.

In Definition \ref{def:localization}, we assume that $X$, $Y$, and $Z$ are finite in order to avoid technical complications involving measure theoretic formulation. However, in Lemma \ref{lem:closeness_localizable_POVM}, we show that if 
$Z$ is finite, then there is no loss in generality in assuming that the sets $X$ and $Y$ are finite.

It is natural to guess that the resource state must have a sufficiently large entanglement to localize a POVM.
In fact, since the Schmidt rank does not increase under stochastic local operations and classical communication (LOCC) \cite{Dur_Vidal_Cirac_2000}, neither does the Schmidt number \cite{Terhal_Horodecki_2000SchmidtNumber} (denoted by $\schn$), the extension of the Schmidt rank to mixed states.
We must have
\begin{eqnarray}
    \label{eq:schmidt_number1}  \schn ( \refs ) \geq \max_{c \in Z} ~ \schn ( M_c ),
\end{eqnarray}
by considering the bi-partition between Alice and Bob.
We also have
\begin{eqnarray}
    \label{eq:schmidt_number2}    \max_{a \in X} ~ \schn (A_a),~\max_{b \in Y} ~ \schn (B_b) \geq \max_{c \in Z} ~ \schn ( M_c ),
\end{eqnarray}
by the same reasoning applied to the bi-partitions $\sysa \otimes  (\refa \otimes \sysb \otimes \refb)$ and $\sysb \otimes  (\refb \otimes \sysa \otimes \refa)$.
However, we have to keep in mind that the classical communication is not allowed for our state manipulation.
Unlike the case of LOCC, there are certain partially entangled states that cannot be obtained from maximally entangled states under this communication restriction \cite{Schmid_2023understanding}. Therefore, it remains unclear whether maximally entangled states are the most powerful resource for localization among states with a fixed Schmidt number.

Early studies of localizable POVMs, which were sometimes referred to as ``instantaneous measurements,'' offered iterative protocols for localizing any bipartite POVM, provided unlimited entanglement resources were available \cite{Groisman_2001,Groisman_2003optics,Vaidman_2003}.
Efforts to limit the consumption of entanglement followed, including a protocol consuming finite entanglement on average but still requiring unlimited entanglement in the worst case \cite{Clark_2010}.
The blind-teleportation protocol of \cite{Vaidman_2003} was refined by Pauwels \emph{et al.} \cite{pauwels_2025classification} so that some POVMs require only finite iteration rounds and entanglement.
They identified all two-qubit PVM measurements that can be localized using the refined blind-teleportation protocol with $1$-ebit and $3$-ebit resources.

\subsection{Linear operators and bipartite vectors}
A POVM is referred to as rank-1 if all its elements are rank-1 operators.
If no pair of POVM elements are proportional to each other, the POVM is said to be \emph{non-redundant}.
A localization is referred to as rank-1 and non-redundant if it comprises rank-1 and non-redundant POVMs, respectively.

We employ the ``double--ket'' notation, also used in e.g. Chiribella \emph{et al.} \cite{chiribella_theoretical_2009}, to represent bipartite vectors by linear operators.
For the Hilbert space $\hil$, we define a bipartite vector $\dket[\id_\hil] = \sum_i \ket[i] \ket[i]$ with a fixed computational basis $\{ \ket[i] \}_{i=1,\ldots,\dim \hil}$.
Let $\hil_\mathrm{in}$ and $\hil_\mathrm{out}$ be Hilbert spaces.
For a linear operator $E:\hil_\mathrm{in} \rightarrow \hil_\mathrm{out}$, we define a bipartite vector $\dket[E]$ on $\hil_\mathrm{in} \otimes \hil_\mathrm{out}$ by
\begin{eqnarray}
    \dket[E] := (\id_{\hil_\mathrm{in}} \otimes E) \dket[\id_{\hil_\mathrm{in}}]. 
\end{eqnarray}
In turn, any bipartite vector can be represented as a double--ket vector of a linear operator.
For example, the two-qudit Schmidt decomposed state $\sum_{i=1}^d \lambda_i \ket[i] \ket[i]$ is represented by the operator $\sum_i \lambda_i \ket[i] \bra[i]$ diagonalized in the computational basis. The application of local linear maps $F_\mathrm{in}:\hil_\mathrm{in} \rightarrow \hil_\mathrm{in}$ and $F_\mathrm{out}:\hil_\mathrm{out} \rightarrow \hil_\mathrm{out}$ to the bipartite vector $\dket[E]$ corresponds to right- and left-multiplication on the operator $E$, respectively, as given by
\begin{equation*}
    F_\mathrm{in} \otimes F_\mathrm{out} \dket[E] = \dket[F_\mathrm{out} E F_\mathrm{in}^\top ],
\end{equation*}
where $\top$  denotes the transpose with respect to the computational basis of $\hil_\mathrm{in}$. This relation illustrates how the Schmidt decomposition of $\dket[E]$ reduces to the singular value decomposition of $E$.
It is a standard fact that the Schmidt rank of vector $\dket[E]$ is equal to the rank of $E$.
We also use a shorthand
\begin{eqnarray}
    \op[E] := \dket[E] \dbra[E],
\end{eqnarray}
to represent rank-1 bipartite operators by linear operators. 

We need to specify the input and output spaces of the linear operator when using the double--ket notation because the definition is not symmetric: $(\id_{\hil_\mathrm{in}} \otimes E) \dket[\id_{\hil_\mathrm{in}}] = (E^\top \otimes \id_{\hil_\mathrm{out}}) \dket[\id_{\hil_\mathrm{out}}]$ where $\top$ is the transpose in the computational basis.
In this article, the double--ket notation is applied to bipartite systems of Alice ($\sysa \otimes \refa$), Bob ($\sysb \otimes \refb$), target POVMs ($\sysa \otimes \sysb$), and resource states ($\refa \otimes \refb$).
We take the input space $\hil_\mathrm{in}$ and output space $\hil_\mathrm{out}$ according to Table~\ref{tab:table1}.
\begin{table}[tbp]
\caption{\label{tab:table1} Choice of input and output spaces in the double-ket notation.}
    \begin{ruledtabular}
        \begin{tabular}{lccc}
            \textrm{System}&
            \textrm{Input } $\hil_\mathrm{in}$&
            \textrm{Output } $\hil_\mathrm{out}$&
            \textrm{Example}\\
            \colrule
            \textrm{Alice} & $\refa$ & $\sysa$ & $\dket[A],\op[A]$\\
            \textrm{Bob} & $\sysb$ & $\refb$ & $\dket[B],\op[B]$\\
            \textrm{Target POVM} & $\sysb$ & $\sysa$ & $\dket[M],\op[M]$\\
            \textrm{Resource state} & $\refb$ & $\refa$ & $\dket[R],\op[R]$
        \end{tabular}
    \end{ruledtabular}
\end{table}

\section{Simplifying localizations}

\subsection{\label{sec:generally_applicable}Generally applicable simplification}
While we concentrate on the localization of rank-1 PVMs as in \cite{pauwels_2025classification}, some results are applicable to the broader class of POVMs and are worth presenting with full generality.
Since this article investigates no-go theorems on localizability, it is vital to simplify the localization and narrow the area of search.
Here we present two directions of simplifications that are generally applicable, specifically, the restriction to rank-1 non-redundant localization and the reduction to finite-outcome POVMs.

Alice's and Bob's operations for localization include arbitrary POVMs in definition \ref{def:localization}.
We first show that the restriction on the local POVMs to rank-1 non-redundant ones does not change the definition of localizability.
\begin{lemma}\label{lem:rank-1_non-redundant}
    If POVM $\{ M_c \}_{c \in Z}$ on $\sysa \otimes \sysb$ can be localized by $\refs$ on $\refa \otimes \refb$, there is a non-redundant rank-1 localization with $\refs$.
\end{lemma}
The proof is provided in Appendix~\ref{appx:proof_rank1nonred}.

This lemma justifies the use of double--ket notation for local POVMs.
In Section \ref{sec:localizability_of_rank1_pvm}, we also find that an analysis restricted to rank-1 resource states is essential, rather than considering mixed states.
The double--ket notation becomes particularly powerful when applied to both POVMs and  resource states. 

Second, we can assume that the numbers of outcomes from local POVM measurements are finite.
\begin{lemma}
    \label{lem:closeness_localizable_POVM}
    If a POVM can be localized by $\refs$, there is a non-redundant rank-1 localization with $\refs$ such that the numbers of Alice's and Bob's measurement outcomes $|X|$ and $|Y|$ satisfy
    \begin{eqnarray}
        |X|,|Y| \leq (|Z|-1) (\dim \sysa \otimes \sysb \otimes \refa \otimes \refb)^2 +1.
    \end{eqnarray}
\end{lemma}
The proof is provided in Appendix \ref{lem:bound_measurement_outcome}.

This lemma shows that the set of POVMs on finite dimensional space $\sysa \otimes \sysb$ with a fixed and finite number of outcomes, localized by fixed resource state $\refs$ in finite dimensional space $\refa \otimes \refb$, is closed.
Thus, the definition of localizable POVMs remains unchanged even when
continuous measurement outcomes are allowed for $X$ and $Y$.

This lemma also shows an inherent limitation on the approximation accuracy achievable for approximately localizable POVM measurements. Although our focus in this work is on exactly localizable POVMs, we may also consider a POVM $\{\tilde{M}_c\}_c$ that is close to a exactly localizable POVM $\{M_c\}_c$ with respect to an appropriate distance measure such as the diamond norm. If $\{\tilde{M}_c\}_c$ is not exactly localizable, then there exists an $\epsilon$-ball around $\{\tilde{M}_c\}_c$ in which no POVM is exactly localizable, since the set of localizable POVMs is closed under any norm (in a finite dimensional vector space).
Consequently, for any POVM $\{\tilde{M}_c\}_c$ that is not exactly localizable, there exists a positive number $\epsilon$ that quantifies the fundamental limit on approximation: the approximation error of $\{\tilde{M}_c\}_c$ under non-adaptive LOCC must be strictly greater than $\epsilon(>0)$. This is in sharp contrast to the situation with (adaptive) LOCC~\cite{CLMOW14}.

\subsection{\label{sec:localizability_of_rank1_pvm}Localizability of rank-1 PVMs}
The localizability condition can be rewritten simply when the target POVM is rank-1 PVM.

Due to the following lemma, we can focus on pure resource states.
\begin{lemma}\label{lem:pure}
    If a rank-1 PVM can be localized by $\left( \refs,~ \{ A_a \}_{a \in X},~ \{ B_b \}_{b \in Y},~ p(Z|XY) \right)$ with a mixed resource state $\refs = \sum_i p_i \psi_i$, it can be localized by $\left( \psi_i,~ \{ A_a \}_{a \in X},~ \{ B_b \}_{b \in Y},~ p(Z|XY) \right)$ for every component $\psi_i$.
\end{lemma}
\emph{Proof}.
Let us denote the right-hand side of Eq.~(\ref{eq:localization}) by $E_{c,\refs}$ so that $\refs$ is regarded as a variable.
When $\refs = \sum_i p_i \psi_i$, we have a decomposition $M_c = \sum_i p_i E_{c,\psi_i}$, which implies $E_{c,\psi_i}\propto M_c$ for all $i$ because $M_c$ is rank-1.
This further implies that $\psi_i$ localizes a POVM $\{ r_c M_c \}_{c \in Z}$ with some real numbers $r_i$.
Since $\{ M_c \}_{c \in Z}$ is a PVM, $\{ r_c M_c \}_{c \in Z}$ must be equal to $\{ M_c \}_{c \in Z}$ itself.
$\qed$\\
Consequently, the localizability by pure resource states also determines the value of mixed resource states.

We shall employ the double--ket notation and represent the target POVM by
\[
    \{ \op[M_c] \}_{c \in Z},
\]
with some linear operators $M_c:\sysb \rightarrow \sysa$.
The local POVMs and the resource state are represented similarly if they are rank-1, in accordance with the rule presented in Table~\ref{tab:table1}.
For any triple of bipartite vectors $\dket[A] \in \sysa \otimes \refa$, $\dket[B] \in \sysb \otimes \refb$ and $\dket[R] \in \refa \otimes \refb$, we have
\begin{eqnarray*}
    \trace_{\refa,\refb}[ (\op[A] \otimes \op[B]) (\id_{\sysa \otimes \sysb} \otimes \op[R])] \\
    = \dbra[R] A,B \rangle \rangle \langle \langle A,B \dket[R],\\
    \dbra[R] A,B \rangle \rangle = (A R^\ast B \otimes \id_\sysb ) \dket[\id_\sysb] = \dket[AR^\ast B],
\end{eqnarray*}
and therefore
\begin{eqnarray}
    \label{eq:simplify} \trace_{\refa,\refb}[ (\op[A] \otimes \op[B]) (\id_{\sysa \otimes \sysb} \otimes \op[R])] = \op[AR^\ast B].
\end{eqnarray}

\begin{lemma}\label{lem:rank-1_localizability}
    A rank-1 PVM $\{ \op[M_c] \}_{c \in Z}$ on $\sysa \otimes \sysb$ can be localized by pure state $\op[R]$ on $\refa \otimes \refb$ if and only if there are rank-1 non-redundant POVMs $\{ \op[A_a] \}_{a \in X}$ and $\{ \op[B_b] \}_{b \in Y}$ and a function $f:X \times Y \rightarrow Z$ that satisfy 
    \begin{eqnarray}
        \label{eq:matrix_condition} A_a R^\ast B_b \propto M_{f(a,b)} \qquad (\forall (a,b) \in X \times Y).
    \end{eqnarray}
\end{lemma}
\emph{Proof}.
From Lemma~\ref{lem:rank-1_non-redundant}, the localizability condition is equivalent to the existence of rank-1 non-redundant localization.
By using Eq.~(\ref{eq:simplify}), the reduced necessary and sufficient condition is written as follows: there are rank-1 non-redundant POVMs $\{ \op[A_a] \}_{a \in X}$ and $\{ \op[B_b] \}_{b \in Y}$ and conditional probability $p(Z|XY)$ such that
\begin{eqnarray}
    \forall c \in Z, \quad \op[M_c] = \sum_{a,b} p(c|a,b) \op[A_a R^\ast B_b].
\end{eqnarray} 
This condition implies $A_a R^\ast B_b\propto M_c$ for some $c$, since $\op[M_c]$ is rank-1.
Therefore,  the function $f$ satisfying (\ref{eq:matrix_condition}) can be defined.

Conversely, assume the existence of rank-1 non-redundant POVMs $\{ \op[A_a] \}_{a \in X}$ and $\{ \op[B_b] \}_{b \in Y}$ and the function $f$ that satisfy (\ref{eq:matrix_condition}).
We can define a conditional probability $p(Z|XY)$ by
\begin{eqnarray}
    p(c|a,b) := \left\{ \begin{array}{cc}
           1  &  c = f(a,b) \\
           0  &  \text{otherwise}.
           \end{array} \right.
\end{eqnarray}
The set of operators $\left\{ \sum_{a,b} p(c|a,b) \op[A_a R^\ast B_b] \right\}_{c \in Z}$ thus defined is a POVM on $\sysa \otimes \sysb$,  whose element with index $c$ is proportional to $\op[M_c]$.
However, such a POVM must be $\{ \op[M_c] \}_{c \in Z}$ itself because the elements $\op[M_c]$ do not overlap.
Therefore, $(\op[R],\{ \op[A_a] \}_{a \in X}, \{ \op[B_b] \}_{b \in Y},p(Z|XY))$ is a localization of $\{ \op[M_c] \}_{c \in Z}$.
$\qed$

This theorem justifies the following definition of localization for rank-1 PVMs, in which the conditional probability is replaced by the function $f:X \times Y \rightarrow Z$.
\begin{definition}[rank-1 localization]\label{def:rank1_localization}
    The tuple $(\op[R],\{ \op[A_a] \}_{a \in X}, \{ \op[B_b] \}_{b \in Y},f)$ that meets the condition of Lemma~\ref{lem:rank-1_localizability} is also called a localization of rank-1 PVM $\{ \op[M_c] \}_{c \in Z}$.
    The function $f:X \times Y \rightarrow Z$ is called the \emph{pattern function} of the localization.
\end{definition}

\section{Localizability of rank-1 PVM by entangled state with bounded Schmidt rank}
We consider the case $\dim \sysa = \dim \sysb=d$ for simplicity.
In this section, we further focus on the situation where the Schmidt rank of the resource state is at most $d$.

This situation is not only a limitation of localization to small reference systems but also a meaningful subclass of LOCC.
Given a maximally entangled state of the equal-sized reference system, any POVM measurement can be implemented via teleportation.
Still, one may seek a method to further reduce the classical communication between parties involved.

We show the following lemma to reduce the scenario into that with ``equal-sized'' measuring and resource systems.
\begin{lemma}
    \label{lem:SchmidtNbound_implies_equalsize}
    A rank-1 PVM can be localized by a mixed resource state $\refs$ with a Schmidt number at most $d$ if and only if it can be localized by a pure resource state in $\refa\otimes\refb$ with $\dim \refa = \dim \refb=d$.
\end{lemma}
\emph{Proof}.
Since `if' part is trivial, we show the converse.
Let $(\refs,\{ \op[A_a] \}_{a \in X}, \{ \op[B_b] \}_{b \in Y},p(Z|XY))$ be a localization of the rank-1 PVM.
Since $\schn ( \refs ) \leq d$, we can further assume that $\refs=\sum_ip_i\psi_i$, $\psi_i$ is pure and $\schn ( \psi_i ) \leq d$ for any $i$.
Lemma \ref{lem:pure} implies that $(\psi_i,\{ \op[A_a] \}_{a \in X}, \{ \op[B_b] \}_{b \in Y},p(Z|XY))$ is a localization of the PVM for any $i$.
Since $\schn (\psi_i ) \leq d$, there exist 
$d$-dimensional subspaces $\refa^d$ of $\refa$ and $\refb^d$ of $\refb$ such that
\begin{eqnarray}
    \ket[\psi_i] \in \refa^d \otimes \refb^d.
\end{eqnarray}
Let $V_A:\mathbb{C}^d\rightarrow\refa$ and $V_B:\mathbb{C}^d\rightarrow\refb$ be isometries onto ranges $\refa^d$ and $\refb^d$, respectively.
By a straightforward calculation, we find that $((V_A\otimes V_B)^\dagger\psi_i(V_A\otimes V_B),\{ V_A^\dagger\op[A_a] V_A\}_{a \in X}, \{V_B^\dagger \op[B_b]V_B \}_{b \in Y},p(Z|XY))$ is a localization of the PVM since $V_AV_A^\dagger\otimes V_BV_B^\dagger$ is the Hermitian projector onto $\refa^d\otimes\refb^d$, whose action does not change $\ket[\psi_i]$. This completes the proof.
$\qed$\\



\subsection{\label{sec:nice_error_basis}Nice unitary error basis}
Let us briefly review the unitary and the nice error bases.
A set of $d^2$ unitary operators $\{ U_i \}_{i=1,\ldots,d^2}$ of dimension $d$ is called a unitary error basis when it forms an orthogonal basis of the Hilbert-Schmidt operator space.
This is equivalent to saying that $\{ d^{-1/2}~\dket[U_i] \}_{i=1,\ldots,d^2}$ is a maximally entangled basis.
The correspondence between maximally entangled bases and unitary error bases is one-to-one.
The local unitary (LU)-equivalence of bipartite bases is translated to equivalence of unitary error bases, specifically, two unitary bases $\{ U_i \}$ and $\{ V_i \}$ are said to be equivalent if there exist unitaries $W_1$ and $W_2$ and some unit complex numbers $c_{i}$ such that
\begin{equation}
    V_i = c_i W_1 U_i W_2, \qquad (\forall i).
\end{equation}
A unitary error basis is called a \emph{nice basis} \cite{knill1996non} if it satisfies
\begin{equation}
    U_i U_j = c_{ij} U_{K(i,j)}, \qquad (\forall i,j),
\end{equation}
where $K:d^2 \times d^2 \rightarrow d^2$ is a function and $c_{ij}$ are some unit complex numbers.
Note that the niceness property is not invariant under LU transformations.

We find a simple and computable characterization of unitary error bases that are LU-equivalent to nice bases.
Such bases are directly related to localizable PVMs in the next subsection.
\begin{theorem}\label{thm:equivalence_to_nice}
    The following three statements on unitary error basis $\{ U_i \}_{i=1,\ldots,d^2}$ are equivalent:
    \begin{description} 
        \item[(i)] $\{ U_i \}_{i=1,\ldots,d^2}$ is LU-equivalent to a nice error basis,
        \item[(ii)] $\{ U_j^\dagger U_i \}_{i=1,\ldots,d^2}$ is a nice error basis for some $j$, and
        \item[(iii)] $\{ U_j^\dagger U_i \}_{i=1,\ldots,d^2}$ is a nice error basis for any $j$.
    \end{description}
\end{theorem}
\emph{Proof}.
(iii)$\Rightarrow$(ii)$\Rightarrow$(i) is trivial. We show (i)$\Rightarrow$(iii). 
Let $W_1$ and $W_2$ be unitaries that make $\{ W_1 U_i W_2 \}_{i=1,\ldots,d^2}$ a nice error basis.
The niceness condition leads to the existence of function $K:[1,d^2] \times [1,d^2] \rightarrow [1,d^2]$ such that $W_1 U_i W_2 W_1 U_j W_2 \propto W_1 U_K(i,j)W_2$.
Applying $W_1^\dagger$ from the left and $W_2^\dagger$ from the right, we have
\begin{eqnarray}
    U_i W_2 W_1 U_j \propto U_{K(i,j)},
\end{eqnarray}
for any pair $(i,j)$.
The function $K$ forms a Latin square, namely, both $K(i,-):[1,d^2] \rightarrow [1,d^2]$ and $K(-,j):[1,d^2] \rightarrow [1,d^2]$ are injective for all $i,j$, because $U_i$ are mutually distinct full-rank matrices and because $U_i W_2 W_1$ and $W_2 W_1 U_j$ are full-rank.
Consequently, function $J:[1,d^2] \times [1,d^2] \rightarrow [1,d^2]$ can be defined by
\begin{eqnarray}
    J(k,i) = j \quad \Leftrightarrow \quad K(i,j) =k,
\end{eqnarray}
and also forms a Latin square.
Using these functions, we have
\begin{eqnarray*}
    U_{i_1} U_j^\dagger U_{i_2} \propto U_{i_1} U_j^\dagger U_j W_1 W_2 U_{J(i_2,j)} = U_{i_1} W_1 W_2 U_{J(i_1,j)} \\
    \propto U_{K(i_1,J(i_2,j))},
\end{eqnarray*}
and thus
\begin{eqnarray}
    U_j^\dagger U_{i_1} U_j^\dagger U_{i_2} \propto U_j^\dagger U_{K(i_1,J(i_2,j))},
\end{eqnarray}
for any pair $(i_1,i_2)$.
$\qed$\\
This theorem implies that the LU-equivalence to a nice error basis can be efficiently verified by checking that for some $l$, $\forall i,\forall j,\exists k,(U_l^\dagger U_i)(U_l^\dagger U_j)\propto U_l^\dagger U_k$ .
As a specific instance, a unitary error basis including identity is LU-equivalent to a nice basis if and only if it is itself a nice basis since we can let $U_l=\id$.

A unitary error basis that is not LU-equivalent to a nice one is called wicked, and its example was shown by Klappenecker and R\"{o}tteler \cite{Klappenecker_2005}.
Theorem \ref{thm:equivalence_to_nice} is useful for finding other examples of wicked bases.
Musto and Vicary \cite{musto2016a} and Beckman \emph{et al.} \cite{Beckman_2001} for example, explicitly constructed unitary error bases that are not themselves nice at $d=4$.
Since those bases include the identity operator $\id_4$, they are wicked by Theorem~\ref{thm:equivalence_to_nice}.

\subsection{rank-1 PVMs with maximal Schmidt rank}
This section focuses particularly on the rank-1 PVMs whose elements have the maximal Schmidt rank, i.e., $d$.
To our surprise, the only localizable PVMs under this constraint are maximally entangled bases generated by nice unitary error bases.

Our first main result directly connects the nice error bases and the localizable rank-1 PVMs.
\begin{theorem}\label{thm:lcalizable=nice}
    Let $\{ \op[M_i] \}_{i=1, \ldots,d^2}$ be a rank-1 PVM on a $d \times d$ dimensional space that has at least one element with the maximal Schmidt rank.
    The PVM can be localized by a $d \times d$ dimensional resource state if and only if it is a maximally entangled basis such that $\{ M_j^{-1} M_i \}_{i=1, \ldots,d^2}$ are nice unitary error bases for $j=1,\ldots,d^2$.
\end{theorem}
A localization of PVMs satisfying the conditions of Theorem \ref{thm:lcalizable=nice} can be explicitly constructed: Alice's PVM $\{ \op[M_i] \}_\iidd$, Bob's PVM $\{ \op[M_i] \}_\iidd$, and the resource state $\op[M_j^{T}]$, where any $j \in [1,d^2]$ works.
The localization conditions of Lemma~\ref{lem:rank-1_localizability} can be easily checked.
In fact, the product $A_a R^\ast B_b$ now reads
\begin{eqnarray}
    M_i M_j^{\dagger} M_k=\frac{1}{d}M_i M_j^{-1} M_k,
\end{eqnarray}
which can be verified to coincide with one of $\frac{1}{d}\{ M_i \}_\iidd$ (up to phase) by the niceness condition.
The full proof of Theorem~\ref{thm:lcalizable=nice} including the ``only if'' part is more involved and is provided in Appendix~\ref{appx:proof_nice}.

Therefore, according to Theorem~\ref{thm:lcalizable=nice}, $\{ \op[M_i] \}_\iidd$ can be localized by an equal-sized resource if and only if it is a maximally entangled basis such that $\{ \sqrt{d} M_i \}_\iidd$ is LU-equivalent to a nice error basis.
We refer to maximally entangled bases satisfying the equivalent conditions of Theorem \ref{thm:equivalence_to_nice} as the ``nice Bell bases.''
Note that, as stated immediately after the proof of Theorem \ref{thm:equivalence_to_nice}, it is possible to efficiently determine whether a given basis is a nice Bell basis.

Our finding reveals the class of localizable rank-1 PVMs with maximal Schmidt rank is strongly limited by equal-sized resources.
The equal-sized resources cannot localize POVMs that have maximal Schmidt rank but are not maximally entangled.
The class screened out by this criterion includes many iso-entangled bases from \cite{delSanto_2024} and partially entangled bases, such as
\[
    \left\{ \ket[00], ~\ket[11], ~\frac{\ket[01] + \ket[10]}{\sqrt{2}},~\frac{\ket[01] - \ket[10]}{\sqrt{2}} \right\}
\]
(named ``pBSM'' in \cite{pauwels_2025classification}).
Among iso-entangled bases, the higher-dimensional generalization \cite{Czartowski_2021bipartitequantum} of the elegant joint measurement \cite{gisin_2019ejm} has the maximal Schmidt rank but is not maximally entangled for any dimension.

Merely being a maximally entangled basis is not enough to be localized by an equal-sized resource.
The basis must also be LU-equivalent to a nice error basis.
Two-qubit POVMs do not suffer additional constraints because any unitary error basis is LU-equivalent to the Pauli basis that is nice.
For higher dimensions, wicked unitary error bases, shown in \cite{Klappenecker_2005} and Section \ref{sec:nice_error_basis}, do not define nice Bell bases.
A maximally entangled basis that cannot be localized by equal-sized resources does exist.

\subsection{Two-qubit POVMs}
We completely characterize two-qubit rank-1 PVMs that can be localized by a resource state with a Schmidt number at most $2$.
Before showing the main theorem, we introduce the LU-equivalence between two bases.
\begin{definition}
    Two bases $\{\ket[\phi_c]\in\sysa\otimes\sysb\}_c$ and $\{\ket[\psi_c]\in\sysa\otimes\sysb\}_c$ are LU-equivalent if there exist LU operators $u_A$ and $u_B$ acting on $\sysa$ and $\sysb$, respectively, such that
    \begin{equation}
        \{(u_A\otimes u_B)\ketbra[\phi_c](u_A\otimes u_B)^\dagger\}_c=\{\ketbra[\psi_c]\}_c.
    \end{equation}
\end{definition}
This definition allows us to identify two bases if the associated POVMs are equivalent as a set of positive semi-definite operators. Note that the localizability of a measurement basis is invariant under the LU-equivalence.
Our second main result is the following:
\begin{theorem}
    \label{thm:2Qlocalizable}
    A two-qubit rank-1 PVM can be localized by a resource state with a Schmidt number at most $2$ if and only if the measurement basis is LU-equivalent to either
    \begin{itemize}
        \item the computational basis, or
        \item the Bell basis, or
        \item the BB84 basis $\{ \ket[00],\ket[01],\ket[1+],\ket[1-] \}$ or $\{ \ket[00], \ket[10], \ket[+1], \ket[-1] \}$.
    \end{itemize}
\end{theorem}
A weaker version of the above theorem was conjectured by Gisin and Del Santo \cite{Gisin_2024towardsmeasurement}, where the resource state is assumed to be a Bell state.
Pauwels \emph{et al.}~\cite{pauwels_2025classification} showed that the theorem holds when one relies on the blind-teleportation protocol, where the BB84 is called $\frac{\pi}{2}$-twisted basis measurement. Consequently, the `if' direction can be shown using their protocol. In contrast, the `only if' direction requires a separate proof, since the theorem permits arbitrary localization protocols beyond the blind teleportation.
We complement their protocol-based analysis by an algebraic approach and thereby completely affirmed the conjecture.

Our analysis is made possible by the simple structure of a two-qubit system: vectors are either product or of full Schmidt rank.
We can use Theorem~\ref{thm:lcalizable=nice} to eliminate all entangled POVMs, except the Bell measurement.
In general, the basis of a two-qubit system that comprises only product vectors is LU-equivalent to either
\begin{eqnarray}
    \{ \ket[00], \ket[01], \ket[1 e_0], \ket[1 e_1] \},
\end{eqnarray}
or its permutation, where $\{ \ket[e_0], \ket[e_1] \}$ is an arbitrary local orthonormal basis. The complete proof is provided in Appendix~\ref{appendix:2Qproof}.

\section{Ideal measurements in nice Bell bases}
So far, we have investigated the localizability of measurements that have only classical outcomes.
The situation changes when Alice and Bob need to perform an ideal measurement \cite{Gisin_2024towardsmeasurement}, that is, to output the projected post-measurement state as well as the classical outcome.
In this section, we apply our analysis to localizing ideal projective measurements in the nice Bell basis.

Ideal projective measurements constitute a specific class of quantum instruments. A quantum instrument (hereafter referred to as an instrument) is characterized by a set of trace-non-increasing completely positive maps $\{ \E_c \}_{c \in Z}$ that sums up to a trace-preserving completely positive map, where $Z$ denotes the index set of classical outcomes. To implement a joint instrument locally, both Alice and Bob must perform local instruments:
\begin{definition}[localization of instruments]\label{def:localization_instrument}
    Let $Z$ be a finite set and $\{ \E_c \}_{c \in Z}$ be an instrument from $\sysa^\mathrm{in} \otimes \sysb^\mathrm{in}$ to $\sysa^\mathrm{out} \otimes \sysb^\mathrm{out}$.
    It is defined to be localized by state $\refs$ on $\refa \otimes \refb$ if there exist instrument $\{ \E_a^A \}_{a \in X}$ from $\sysa^\mathrm{in} \otimes \refa$ to $\sysa^\mathrm{out}$, instrument $\{ \E_b^B \}_{b \in Y}$ from $\sysb^\mathrm{in} \otimes \refb$ to $\sysb^\mathrm{out}$ and a conditional probability $p(Z|XY)$ such that
    \begin{eqnarray}
        \label{eq:localization_ideal}    \E_c( \rho_\mathcal{S} ) = \sum_{a,b} p(c|a,b) \E^A_a \otimes \E^B_b (\rho_\mathcal{S} \otimes \refs)
    \end{eqnarray}
    holds for any state $\rho_\mathcal{S}$ and any $c \in Z$.
\end{definition}

An ideal projective measurement (or just an ``ideal measurement'') on $\sysa \otimes \sysb$ is an  instrument $\{ \E_c \}_{c \in Z}$ from $\sysa \otimes \sysb$ to $\sysa \otimes \sysb$ such that its elements are described by orthogonal projectors $P_c$ on $\sysa \otimes \sysb$ as $\E_c (\rho_\mathcal{S})=P_c \rho_\mathcal{S} P_c$ ($c \in Z$).
It is ``ideal'' in that the measurement can be repeated and produce the same result at all repetitions.
When applied to ideal measurements, Definition \ref{def:localization_instrument} differs slightly from the one in \cite{Beckman_2001}, where the focus of localization is the decoherence map $\sum_{c \in Z} \E_c$.
Localizability of ideal measurements in the sense of Definition \ref{def:localization_instrument} implies that of \cite{Beckman_2001}.

Although any bipartite (non-ideal) POVM is localizable \cite{Vaidman_2003}, the same is not true for ideal measurements.
All localizable and ideal two-qubit measurements must be LU-equivalent to either product or Bell measurements \cite{Popescu_Vaidman_1994}.
For higher-dimensional systems, Beckman \emph{et al.} \cite{Beckman_2001} derived several necessary conditions for the ideal measurements to be localizable.
We can strengthen this condition by explicitly constructing localization protocols.
\begin{theorem}\label{thm:Beckman}
    An ideal measurement on a two-qudit basis that has at least one element with the maximal Schmidt rank can be localized if and only if it is a nice Bell basis.
\end{theorem}
This generalizes the localization of the two-qubit Bell measurement \cite{Popescu_Vaidman_1994,Gisin_2024towardsmeasurement} to a high-dimensional regime.\\
\emph{Proof}.
All localizable measurements in our definition persist to be localizable in the definition of \cite{Beckman_2001}.
Noting this, the necessity of being a maximally entangled basis follows directly from Theorem 3 of \cite{Beckman_2001} and the discussion thereafter.
Theorem 6 of \cite{Beckman_2001} states that if $\dket[\id_d/d]$, $\dket[U/d]$, and $\dket[V/d]$ are all included in the basis of a localizable ideal measurement, where $U,V$ are unitaries, then so is $\dket[UV/d]$.
Taking Theorem~\ref{thm:equivalence_to_nice} and LU transformations into account, we can eliminate the assumption that $\dket[\id_d/d]$ is included and instead simply state the following: if an ideal measurement in a maximally entangled basis can be localized, then it must be a nice Bell basis.

\begin{figure}[b]
\includegraphics[width=.45\textwidth]{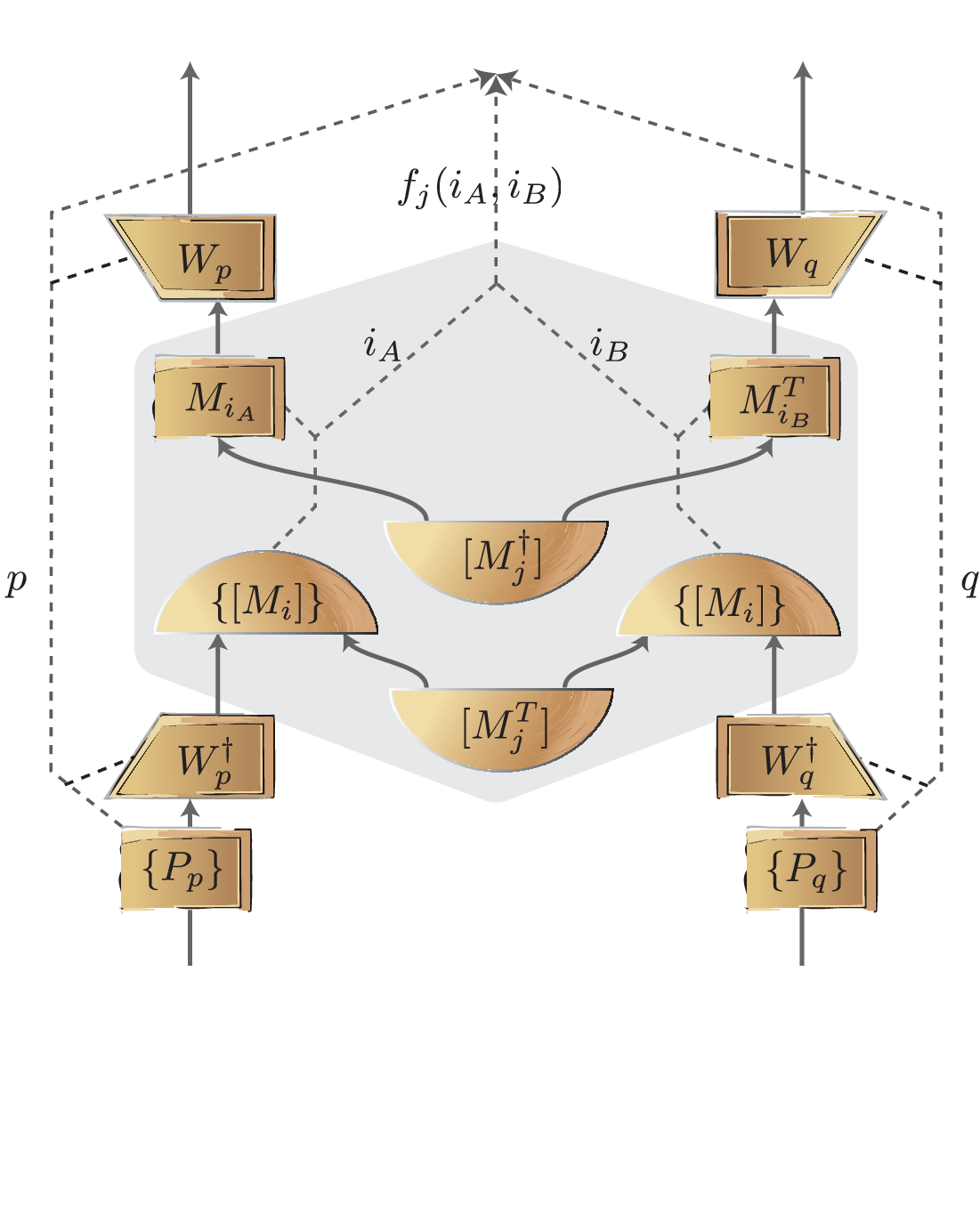}
\caption{\label{fig:localize_ideal_block} The localization scheme for the ideal measurement in the basis (\ref{eq:twisted_nice_Bell}). The shaded area corresponds to the sub-scheme for localizing the ideal measurement in the nice Bell basis $\{ \op[M_i] \}_\iidd$.}
\end{figure}

Now we prove the converse by explicitly building the localization protocol for ideal measurements in nice Bell bases.
Let $\{ \dket[M_i] \}_\iidd$ be the PVM of a nice Bell measurement.
Refer to the shaded area in Fig.~\ref{fig:localize_ideal_block} for clarification.
The resource state is given by
\begin{eqnarray}
    \label{eq:resource} \refs =  \op[M_j^{T}] \otimes \op[M_j^{\dagger}]
\end{eqnarray}
on a $(d^2)^2$-dimensional space, where $j \in [1,d^2]$ is fixed but arbitrary.
Alice and Bob first perform local PVM measurements $\{ \op[M_i] \}_\iidd$ on their target system and the half-resource $\dket[M_j]$, which effectively results in PVM measurement on the state vector
\[
    \dket[M_{i_A} M_j^{\dagger} M_{i_B}] = \frac{1}{d}\dket[M_{f_j(i_A,i_B)}],
\]
upon the measurement results $i_A$ for Alice and $i_B$ for Bob (see Eq.~(\ref{eq:simplify})).
The function $f_j:[1,d^2] \times [1,d^2] \rightarrow [1,d^2]$ is guaranteed to exist by the niceness condition and forms a Latin square.
While Alice does not know $i_B$ and Bod does not know $i_A$,
they can independently apply the unitaries $\sqrt{d}M_{i_A}$ and $\sqrt{d}M_{i_B}^\top$ on the remaining resource state $\dket[M_j^{\dagger}]$ and output the state
\[
    d(M_{i_A} \otimes M_{i_B}^\top) \dket[ M_j^{\dagger} ] = d\dket[M_{i_A} M_j^{\dagger} M_{i_B}] = \dket[M_{f_j(i_A,i_B)}],
\]
together with the results of their measurement $(i_A,i_B)$.
The third person to receive the result $(i_A,i_B)$ knows the measurement result $f_j(i_A,i_B)$.
$\qed$

Building upon the nice Bell bases, we can generalize other examples of localizable ideal measurements provided in \cite{Beckman_2001}\footnote{While a different definition of localizability is used in \cite{Beckman_2001}, their example persists to be localizable in our criterion as well.}.
Let $\sysa = \oplus_{p=1}^{r_A} \sysa^p$ and $\sysb = \oplus_{q=1}^{r_B} \sysb^q$, where the subspaces $\sysa^p$
and $\sysb^q$ all have dimension $d$.
Let $\{ \dket[M_i] \}_\iidd$ be a nice Bell basis on $\sysa^1 \otimes \sysb^1$, $W^A_p:\sysa^1 \rightarrow \sysa^p$ ($p=1,\ldots,r_A$) and $W^B_q:\sysb^1 \rightarrow \sysb^q$ ($q=1,\ldots,r_B$) all be unitary isomorphisms between subspaces.
We can see that
\begin{eqnarray}
    \label{eq:twisted_nice_Bell}    \{ W^A_p \otimes W^B_q \dket[M_i] \}_{\iidd; ~p=1,\ldots,r_A;~ q=1,\ldots,r_B},
\end{eqnarray}
is an orthonormal basis of $\sysa \otimes \sysb$, including the case $d=1$ in which this reduces to a product basis.
The example of a localizable ideal measurement presented in \cite{Beckman_2001} corresponds to the case $d=2$ and where all $W$s are identity isomorphisms.

The localization protocol for the basis (\ref{eq:twisted_nice_Bell}) is divided into three steps.
Refer to Fig.~\ref{fig:localize_ideal_block} for clarification.
The reference system has $d^2 \times d^2$ dimension, and the resource state is the same as Eq.~(\ref{eq:resource}).
First, Alice and Bob perform the ``which subspace'' ideal measurement on their local subsystems, described by the projectors $P_p^A$ on $\sysa^p$ $(p=1,\ldots,r_A)$ and $P_q^B$ on $\sysb^q$ $(q=1,\ldots,r_B)$.
Subsequently, they perform unitaries $W_p^{A \dagger}$ and $W_q^{B \dagger}$, respectively, on the basis of the obtained results $p$ and $q$.
Second, they localize the ideal measurement on the nice Bell basis $\{ \dket[M_i] \}_\iidd$ on $\sysa^1 \otimes \sysb^1$, using the resource state.
When the result is $(i_A,i_B)$, they have effectively performed a non-ideal PVM measurement on the basis (\ref{eq:twisted_nice_Bell}) and produced a state $\dket[M_{f_j(i_A,i_B)}]$ on $\sysa^1 \otimes \sysb^1$ up to here, while individually they are not aware of the value $f_j(i_A,i_B)$.
Third, they independently apply unitaries $W_p^A$ and $W_q^B$ on the basis on the results $p$ and $q$ of the first measurement.
This completes the localization of the ideal measurement.

Now that measurements in bases (\ref{eq:twisted_nice_Bell}) are ideally localizable, one may ask if this is the only such kind.
In fact, Beckman \emph{et al.} \cite{Beckman_2001} showed that the basis must be maximally entangled within partitioned subspaces and that the maximally entangled vectors in different partitions must be related in a certain manner.
We believe that the following conjecture is true:
\begin{conjecture}
    A bipartite rank-1 ideal measurement can be localized if and only if the measurement basis is LU-equivalent to (\ref{eq:twisted_nice_Bell}).
\end{conjecture}

\section{Discussion}
In this work, we investigated bipartite PVMs that can be localized by resource entangled states with a Schmidt numbers not exceeding their local dimensions. In contrast to the previous study by Pauwels \emph{et al.} \cite{pauwels_2025classification} on localizable PVMs with finite entanglement, our approach focuses on protocol-independent algebraic constraints, thereby enabling non-localizable PVMs to be detected.

We demonstrated that a rank-1 PVM on $\mathbb{C}^d \otimes \mathbb{C}^d$ containing at least one
element with the maximal Schmidt rank $d$ can be localized by an entangled state with a Schmidt number at most $d$ if and only if it forms a maximally entangled basis corresponding to a nice Bell basis. Considering that standard quantum teleportation enables the entanglement-assisted implementation of such PVMs via one-way communication, our result reveals a stringent restriction imposed by the non-adaptivity of local operations. In establishing this necessary and sufficient condition, we derived a simple characterization of nice unitary error bases, which further led to new examples of ``wicked" unitary error bases \cite{Klappenecker_2005} being discovered.

For two-qubit systems, we analyzed localizable PVMs without any assumptions on the Schmidt rank. We found that a two-qubit rank-1 PVM can be localized using two-qubit entanglement if and only if it is LU-equivalent to a product basis, a Bell basis, or a BB84-basis measurement. This affirmatively resolves the conjecture by Gisin and Del Santo \cite{Gisin_2024towardsmeasurement} regarding two-qubit localizable PVMs, which had remained open in the work of Pauwels \emph{et al.} \cite{pauwels_2025classification}.

As an application of our analysis, we constructed localization protocols for ideal measurements on nice Bell bases. Specifically, these protocols, combined with the existing results \cite{Beckman_2001}, reveal that an ideal measurement on a two-qudit basis with at least one maximal Schmidt rank element can be localized if and only if it forms a nice Bell basis. While these protocols provide only a sufficient condition for localizable ideal measurements without assuming a specific Schmidt rank, we conjecture that this condition is also necessary.

The analysis of PVMs may become considerably more complex if the assumptions on the Schmidt numbers of the PVMs and resource states are relaxed. For instance, the requirement that pattern functions form Latin squares---a crucial element in the proof of Theorem~\ref{thm:lcalizable=nice}---is no longer mandatory without these assumptions. Our complete characterization of two-qubit localizable PVMs was facilitated by the exceptionally simple structure of $\mathbb{C}^2 \otimes \mathbb{C}^2$, where any bipartite vector either has maximal Schmidt rank ($d=2$) or is a product vector.

Nevertheless, the fundamental findings presented in Sections \ref{sec:generally_applicable} and \ref{sec:localizability_of_rank1_pvm} remain valid in general settings. In particular, we believe that our algebraic formalism based on double--ket notation will be equally effective in higher dimensions. This formalism provides a representation of PVMs and resource states where the localizability condition reduces to a simple matrix equation (see Definition \ref{def:rank1_localization}). Our results rely heavily on matrix analysis within this notation, particularly through the use of matrix inversion, conjugation, and transposition.

More broadly, the analysis of localizable measurements can be viewed as the investigation of fundamental limitations on the power of local operations and shared randomness (LOSR) for implementing joint quantum operations. LOSR operations have attracted significant attention in the contexts of nonlocal games~\cite{Bell_PhysicsPhysiqueFizika.1.195,Bell_RevModPhys.86.419}, self-testing~\cite{MayersY04,Supic2020selftestingof}, semi-quantum nonlocal games~\cite{Buscemi12}, and quantum resource theories~\cite{Schmid_2023understanding,Zjawin2023quantifyingepr}. From this viewpoint, our algebraic approach is expected to have applications beyond the specific setting considered here, potentially contributing to a deeper understanding of quantum nonlocality and resource interconversions (e.g., classical communication versus entanglement) in distributed quantum information processing.

\begin{acknowledgments}
We thank Hiroyuki Osaka for the helpful discussions. We are grateful to Jef Pauwels for reviewing an earlier draft of this manuscript.
This work was supported by Japan Science and Technology Agency (JST) as part of Adopting Sustainable Partnerships for Innovative Research Ecosystem (ASPIRE), Grant no.JPMJAP25A3, and JST CREST, Grant no.JPMJCR25I5.
S.A. was partially supported by JST PRESTO Grant no.JPMJPR2111, JST Moonshot R\&D MILLENNIA Program (Grant no.JPMJMS2061), JPMXS0120319794, and CREST (Japan Science and Technology Agency) Grant no.JPMJCR2113.
J.M. was supported by JSPS KAKENHI Grant no.JP23K21643.
\end{acknowledgments}


\bibliography{reference}

\appendix

\section{\label{appx:proof_rank1nonred}Proof of Lemma~\ref{lem:rank-1_non-redundant}}
In this appendix, we do not employ the double--ket notation as we also deal with degenerate operators $M_c$, $A_a$, $B_b$ and $\refs$.

We prove Lemma~\ref{lem:rank-1_non-redundant} by explicitly constructing the non-redundant rank-1 localization from an arbitrary localization.

Let $\left( \refs,~ \{ A_a \}_{a \in X},~ \{ B_b \}_{b \in Y},~ p(Z|XY) \right)$ be a localization of $\{ M_c \}_{c \in Z}$.
In the first step, we construct rank-$1$ POVMs $\{ A'_{a'} \}_{a' \in X' }$ and $\{ B'_{b'} \}_{b' \in Y' }$ from $\{ A_a \}_{a \in X}$ and $\{ B_b \}_{b \in Y}$.
In short, the refinements of POVMs by spectral decomposition do the job.

Let $A_a = \sum_{i \in [1, \mathrm{rank} A_a]} r_i P_{a,i}$ be a spectral decomposition of $A_a$ (here $[1,N]$ is the shorthand for the set $\{1,\ldots,N \}$ for the natural number $N$) and define
\[
    A'_{a,i} := r_i P_{a,i}, 
\]
for $a \in X$ and $i \in [1, \mathrm{rank} A_a]$.
Define a new index set $X'$ as $\cup_{a \in X} \{ a \} \times [1, \mathrm{rank} A_a]$.
Then the set $\{ A'_{a,i} \}_{(a,i) \in X'}$ of rank-$1$ positive operators is a POVM since $\{ A_a \}_{a \in X}$ is.
Similarly, we define the index set $Y'$ and a rank-$1$ POVM $\{ B'_{a,i} \}_{(b,i) \in Y'}$ from $\{ B_b \}_{b \in Y}$.
The conditional probability $p'(Z|X'Y')$ is defined by
\[
    p'(c|(a,i),(b,j)) := p(c|a,b), \quad (\forall (a,i) \in X', ~(b,j) \in Y').
\]
Then we can verify that the tuple $(\refs,\{ A'_{a,i} \}_{(a,i)}, \{ B'_{b,j} \}_{(b,j)},p' )$ is a rank-1 localization of $\{ M_c \}_{c \in Z}$ by a straightforward calculation.

For the sake of brevity, we redefine the localization $(\refs,\{ A_a \}_{a \in X}, \{ B_b \}_{b \in Y},p )$ to be a rank-1 localization, which is shown to exist from the above argument, and now construct a non-redundant localization while keeping the rank-1 property.

We introduce an equivalence relation $\sim$ in the set $X$ by $a_1 \sim a_2$ iff $A_{a_1} \propto A_{a_2}$.
The elements of the quotient space $X/\sim$ are represented as $[a] = \{ a_X \in X | a_X \sim a \}$ using elements $a$ of $X$.
The equivalence relation and the quotient space are introduced to the index set $Y$ in the same manner.

Now define non-redundant rank-1 POVMs $\{ A'_{a'} \}_{a' \in X/\sim}$ and $\{ B'_{b'} \}_{b' \in Y/\sim}$ by
\[
    A'_{[a]} := \sum_{a_X \in [a]} A_{a_X}, \quad B'_{[b]} := \sum_{b_Y \in [b]} B_{b_Y},
\]
and the conditional probability $p'(Z|X/\sim,Y/\sim)$ by
\[
    p'(c|[a],[b]):=\sum_{a_X \in [a]}\sum_{b_Y \in [b]}\frac{\trace[A_{a_X}]}{\trace[A'_{[a]}]}\frac{\trace[B_{b_Y}]}{\trace[B'_{[b]}]}p(c|a_X,b_Y).
\]
Again, we can verify that $(\refs,\{ A'_{a'} \}_{a' \in X/\sim},\{ B'_{b'} \}_{b' \in Y/\sim},p')$ is a localization of $\{ M_c \}_{c \in Z}$ by a straightforward calculation.

$\qed$

\section{Proof of Lemma~\ref{lem:closeness_localizable_POVM}}
\label{lem:bound_measurement_outcome}

In this section, we prove Lemma~\ref{lem:closeness_localizable_POVM}, which provides bounds on the number of measurement outcomes in localization protocols.
The set of linear operators acting on $\mathcal{H}$ is denoted by $\linop[\hil]$.

\emph{Proof}.
    Let $\left( \refs,~ \{ A_a \}_{a \in X},~ \{ B_b \}_{b \in Y},~ p(Z|XY) \right)$ be a localization of a POVM $\{M_c\}_{c\in Z}$.
    By using Lemma~\ref{lem:rank-1_non-redundant}, we can assume the POVMs are rank-1 and non-redundant.
    Let $\Omega=\sum_{a\in X}A_a\otimes C^{(a)}$, where $C^{(a)}=\sum_{b\in Y,~c\in Z}p(c|a,b)B_{b}\otimes\ketbra[c]$ is the Choi operator of a completely positive and trace preserving (CPTP) map $\linop[\sysb\otimes\refb]\rightarrow \linop[\mathbb{C}^{|Z|}]$.
    Define $\mathcal{A}:=\{A_a\otimes C^{(a)}\}_{a\in X}$, which can be regarded as a subset of $\mathbb{R}^{D}$ with $D=(|Z|-1)(\dim\sysa\dim\refa\dim\refb\dim\sysb)^2$ due to the trace preserving constraint on $C^{(a)}$. Then, we find that $\frac{1}{|X|}\Omega\in conv(\mathcal{A})$.
    By using the Carath\'{e}odory's theorem, there exist $\hat{X}\subseteq X$ and probability distribution $q(\hat{X})$ such that $|\hat{X}|\leq D+1$ and
    \begin{equation}
        \frac{1}{|X|}\Omega=\sum_{a'\in\hat{X}}q(a')A_{a'}\otimes C^{(a')}.
    \end{equation}
    Since 
    \begin{eqnarray}
        &&|X|\sum_{a'\in\hat{X}}q(a')A_{a'}\\
        &=&|X|\sum_{a'\in\hat{X}}q(a')A_{a'}\trace_{ \sysb\otimes\refb\otimes\mathbb{C}^{|Z|}}[\rho_BC^{(a')}]\\
        &=&\trace_{ \sysb\otimes\refb\otimes\mathbb{C}^{|Z|}}[\rho_B\Omega]=\sum_{a\in X}A_a=\id
    \end{eqnarray}
    for any state $\rho_B$, we find that $\{\hat{A}_{a'}:=|X|q(a')A_{a'}\}_{a'\in \hat{X}}$ is a valid POVM.
    Moreover, we can verify $(\refs,\{\hat{A}_{a'}\}_{a'\in\hat{X}},\{B_b\}_{b\in Y},p(Z|\hat{X},Y))$ is a localization of $\{M_c\}_{c\in Z}$ as follows, where $p(Z|\hat{X},Y)$ is defined by restricting $X$ in $p(Z|X,Y)$ into $\hat{X}$.
    \begin{widetext}
    \begin{eqnarray}
        & \sum_{a' \in \hat{X},~ b \in Y} p(c|a',b) \trace_{\refa,\refb} \left[ (\hat{A}_{a'} \otimes B_{b} )~ (\id_{\sysa \otimes \sysb} \otimes \refs ) \right] \\
        =& \sum_{a' \in \hat{X},~ b \in Y} |X|p(c|a',b)q(a') \trace_{\refa,\refb} \left[ (A_{a'} \otimes B_{b} )~ (\id_{\sysa \otimes \sysb} \otimes \refs ) \right] \\
        =& \sum_{a' \in \hat{X}} |X|q(a') \trace_{\refa,\refb} \left[ (A_{a'} \otimes \bra[c]C^{(a')}\ket[c] )~ (\id_{\sysa \otimes \sysb} \otimes \refs ) \right] \\
        =& \trace_{\refa,\refb} \left[ \bra[c]\Omega\ket[c]~ (\id_{\sysa \otimes \sysb} \otimes \refs ) \right] \\
        =& \sum_{a \in X}  \trace_{\refa,\refb} \left[ (A_{a} \otimes \bra[c]C^{(a)}\ket[c] )~ (\id_{\sysa \otimes \sysb} \otimes \refs ) \right]\\
        =& \sum_{a \in X,~ b \in Y} p(c|a,b) \trace_{\refa,\refb} \left[ (A_{a} \otimes B_{b} )~ (\id_{\sysa \otimes \sysb} \otimes \refs ) \right]=M_c.
    \end{eqnarray}
    \end{widetext}
    By construction, $\{\hat{A}_{a'}\}_{a'\in\hat{X}}$ is a rank-1 and non-redundant POVM and $|\hat{X}|\leq D+1$.
    By using the same argument, we can show the lemma.
$\qed$

\section{\label{appx:proof_nice}Proof of Theorem~\ref{thm:lcalizable=nice}}
In this appendix, it is assumed that all subsystems share the same dimension $d$ and that at least one element of the target PVM $\{ \op[M_i] \}_{i=1,\ldots,d^2}$, say $\op[M_1]$, has the maximal Schmidt rank.

\begin{lemma}\label{lem:fullrank_latin}
    Let $(\op[R],\{ \op[A_a] \}_{a \in X}, \{ \op[B_b] \}_{b \in Y},f)$ be a rank-1 non-redundant localization of $\{ \op[M_i] \}_{i=1,\ldots,d^2}$.
    The following three hold:
    \begin{itemize}
        \item The rank-1 POVMs $\{ \op[A_a] \}_{a \in X}$ and $\{ \op[B_b] \}_{b \in Y}$ must be PVMs ($|X| = |Y| =d^2$).
        \item All elements of $\{ M_i \}_\iidd$, $\{ A_a \}_{a \in X}$ and $\{ B_b \}_{b \in Y}$ must be of full-rank.
        \item The pattern function $f$ forms a Latin square, that is, functions $f(a,-):Y \rightarrow [1,d^2]$ and $f(-,b):X \rightarrow [1,d^2]$ are injective for all $a \in X$ and $b \in Y$.
    \end{itemize}
\end{lemma}
\emph{Proof}.
There exists a pair $(a_1,b_1) \in f^{-1}(1)$ satisfying $A_{a_1} R^\ast B_{b_1} = \alpha M_1$ with non-zero $\alpha$, since $\op[M_1] = \sum_{(a,b) \in f^{-1}(1)} \op[A_a R^\ast B_b]$ can never be zero.
Since $M_1$ has the full-rank, so do $A_{a_1}$, $B_{b_1}$, and $R$.

Because $\{ \op[B_b] \}_{b \in Y}$ is non-redundant and $A_{a_1} R^\ast$ has full-rank, $A_{a_1} R^\ast B_b \propto A_{a_1} R^\ast B_{b'}$ implies $b=b'$ and $A_{a_1} R^\ast B_b$ never becomes zero for any $b \in Y$.
In other words, $f(a_1,-):Y \rightarrow [1,d^2]$ is an injective function, which implies that $|Y| \leq d^2$.
Due to the completeness of the rank-1 POVM, we find that $\{ \op[B_b] \}_{b \in Y}$ is a rank-1 PVM with $|Y|=d^2$.
Using the same argument, we find that $\{ \op[B_b] \}_{b \in Y}$ is also a rank-1 PVM with $|X|=d^2$.

Suppose that $B_{b_2}$ is not full-rank.
In this case, $A_a R^\ast B_{b_2}$ can never be proportional to the full-rank matrix $M_1$ for any $a$, which implies that
\begin{eqnarray}
    \trace[ M_1^\dagger A_a R^\ast B_{b_2} ] = 0 \qquad (\forall a \in X).
\end{eqnarray}
However, since $\{ A_a \}_{a \in X}$ forms an operator basis, this implies $R^\ast B_{b_2} M_1^\dagger =0$ and hence $B_{b_2}=0$.
This is impossible since $\{ \op[B_b] \}_{b \in Y}$ is non-redundant, and by contradiction, $B_b$ must be of full-rank for any $b \in Y$.
The same argument shows that $A_a$ must be full-rank for any $a \in X$.

Now that $\{ A_a \}_{a \in X}$, $\{ B_b \}_{b \in Y}$ and $R^\ast$ all have full-rank $d$, so do their products $A_a R^\ast B_b$.
We therefore conclude that $M_i$ has the full-rank for any $\iidd$.
The statement on the pattern function $f$ can be deduced by replacing $M_1$ with $M_i$ ($\iidd$) in the above argument.
$\qed$

\begin{lemma}\label{lem:triple_product}
    There exists a $d \times d$ dimensional resource state that localizes $\{ \op[M_i] \}_\iidd$ if and only if for any $i,j,k \in [1,d^2]$, there is $l \in [1,d^2]$ such that
    \begin{eqnarray}
        \label{eq:triple_product}    M_l \propto M_i M_j^{-1} M_k.
    \end{eqnarray}
\end{lemma}
\emph{Proof}.
If this condition is met, $\{ M_i \}_\iidd$ is localized by state $\psi \propto (M_j^{-1})^\ast$, where $j \in [1,d^2]$ is arbitrary.
If we choose to use $\psi_1 \propto (M_1^{-1})^\ast$ for the resource, the pattern function $f(i,k)=l$ is defined by $M_l \propto M_i M_1^{-1} M_k$.
Then $(\psi_1, \{ \op[M_i] \}_\iidd, \{ \op[M_i] \}_\iidd, f)$ is a localization of $\{ \op[M_i] \}_\iidd$.

Conversely, if $\{ \op[M_i] \}_\iidd$ can be localized, we can assume that the resource is a pure state, say $\op[R]$, without loss of generality (Lemma~\ref{lem:pure}).
This state must have the maximal Schmidt rank from Eq.~(\ref{eq:schmidt_number1}).
From Lemma~\ref{lem:fullrank_latin}, there is a rank-1 non-redundant localization $(\op[R],\{ \op[A_i] \}_\iidd, \{ \op[B_i] \}_\iidd,f)$ such that $A_i$ and $B_i$ are both full-rank.
Since the pattern function $f:[1,d^2] \times [1,d^2] \rightarrow [1,d^2]$ forms a Latin square,
\begin{eqnarray}
    f(i_1,i_2)=i_3 \Leftrightarrow f_1(i_2,i_3)=i_1,~f_2(i_3,i_1)=i_2
\end{eqnarray}
defines two other functions $f_1,f_2:[1,d^2] \times [1,d^2] \rightarrow [1,d^2]$ each forming a Latin square.
Given $j$, take any pair $(j_1,j_2)$ such that $f(j_1,j_2) =j$.
Since $M_j$ is invertible, we have 
\begin{eqnarray}
    M_i M_j^{-1} M_k \\
    \propto A_{f_1(j_2,i)}R^\ast B_{j_2} (A_{j_1} R^\ast B_{j_2})^{-1} A_{j_1} R^\ast B_{f_2(k,j_1)} \\
    = A_{f_1(j_2,i)} R^\ast B_{f_2(k,j_1)} \\
    \propto M_{f(f_1(j_2,i),f_2(k,j_1))},
\end{eqnarray}
as required.
$\qed$

\begin{lemma}\label{lem:must_be_maxent}
    If $\{ \op[M_i] \}_\iidd$ can be localized with a $d \times d$ dimensional state, then $\op[M_i]$ is maximally entangled for all $i$.
\end{lemma}
\emph{Proof}.
The relation (\ref{eq:triple_product}) holds for $\{ \op[M_i] \}_\iidd$ by Lemma~\ref{lem:triple_product}.
If $l=k$ in Eq.~(\ref{eq:triple_product}), we have $\id_d \propto M_i M_j^{-1}$ and thus $i =j$.
By contraposition, we have $l \neq k$ if $i \neq j$ in $M_l \propto M_i M_j^{-1} M_k$.
We find that for any $n$
\begin{eqnarray}
    \label{eq:zero_constraints}  \trace[M_i M_j^{-1} M_k M_k^\dagger] = \trace[M_l M_k^\dagger] = \langle \langle M_k | M_l \rangle \rangle =0,
\end{eqnarray}
holds whenever $i \neq j$.
Since $\{ M_i M_j^{-1} \}_\iidd$ is a (possibly non-orthogonal) basis of the set of $d \times d$ matrices, the $d^2-1$ constraints of (\ref{eq:zero_constraints}) for $i\neq j$
uniquely determine the operator $M_k M_k^\dagger$ up to scalar multiplication.
We also have
\begin{eqnarray}
    M_k M_k^\dagger = M_k \trace_\sysb \left[ \dket[\id_\sysb] \dbra[\id_\sysb] \right] M_k^\dagger = \trace_\sysb \left[ \op[M_k] \right].
\end{eqnarray}
Therefore, there is a density operator $\rho_A$ on $\sysa$ such that
\[
    \trace_\sysb [\op[M_k]] = \rho_A \qquad (\forall k \in [1,d^2]).
\]
A similar argument starting from the transpose of the relation (\ref{eq:triple_product}) leads to the existence of a positive semidefinite operator $\rho_B$ on $\sysb$ such that
\[
    \trace_\sysa [\op[M_k]] = \rho_B \qquad (\forall k \in [1,d^2]).
\]

Since $\dket[M_k]$ are all purifications of $\rho_A$, they have the expression
\begin{eqnarray}
    \label{eq:purification} \dket[M_k] =  (\sqrt{\rho_A} \otimes U_k) \dket[\id_d],
\end{eqnarray}
with some unitary operators $U_k$ on $\sysb$.
This implies
\[
    \rho_B = U_k \sqrt{\rho_A}^\top \trace_\sysa[\op[\id_d]]\sqrt{\rho_A}^\ast U_k^\dagger = U_k \rho_A^\top U_k^\dagger,
\]
and thus
\[
    \rho_B U_k = U_k \rho_A^\top,
\]
holds for any $k$.
Since $\{ U_k \}_{k=1,\ldots,d^2}$ spans the set of linear operators due to the completeness of $\{ M_k \}_{k=1,\ldots,d^2}$, we find that
\[
    \rho_B X = X \rho_A^\top,
\]
holds for any linear operator $X$ on the $d$-dimensional system.
This holds if and only if $\rho_A = \rho_B = \id_d/d$.
From the expression (\ref{eq:purification}), we conclude that $\{ \dket[M_k] \}_{k=1,\ldots,d^2}$ is a maximally entangled basis.
$\qed$\\

\emph{Proof of Theorem~\ref{thm:lcalizable=nice}}.
Combining Lemmas~\ref{lem:triple_product} and \ref{lem:must_be_maxent}, we see that $\{ \op[M_i] \}_\iidd$ can be localized by a $d \times d$ dimensional resource if and only if it is a maximally entangled basis satisfying Eq.~(\ref{eq:triple_product}).
Since $M_j^{-1}$ is full-rank for any $j$, the condition posed by Eq.~(\ref{eq:triple_product}) is equivalent to
\begin{eqnarray}
    \nonumber \forall i,j,k\in [1,d^2],~ \exists l \in [1,d^2] \quad \mathrm{s.t.}\\
    M_j^{-1} M_l \propto M_j^{-1} M_i M_j^{-1} M_k.
\end{eqnarray}
Since $\{ M_{i_1}^{-1} M_{i_2} \}_{i_2 = 1,\ldots,d^2}$ is a unitary error basis for any $i_1$, the above condition is equivalent to the niceness of $\{ M_j^{-1} M_i \}_\iidd$ for any $j$.
$\qed$

\section{\label{appx:proof_2Q}Proof of Theorem~\ref{thm:2Qlocalizable}}
\label{appendix:2Qproof}

In this section, we prove Theorem~\ref{thm:2Qlocalizable}, which completely characterizes the two-qubit localizable rank-1 PVMs.
The set of linear operators acting on $\mathcal{H}$ is denoted by $\linop[\hil]$.

\emph{Proof}.
Lemma~\ref{lem:SchmidtNbound_implies_equalsize} implies that the localizability by a resource state with a Schmidt number at most 2 can be reduced into that by a resouce state in a two-qubit system.
Let $\{\op[M_i]\}_{i=1,\cdots,4}$ be a two-qubit rank-1 PVM localizable by a two-qubit resource state $\dket[R]$.

If $\{\dket[M_i]\}_{i=1,\cdots,4}$ contains an entangled state, Theorem~\ref{thm:lcalizable=nice} implies that it is a maximally entangled basis, i.e., $\{\sqrt{2}M_i\}_{i=1,\cdots,4}$ is a unitary error basis. Since any unitary error basis is LU-equivalent to the Pauli basis that is nice \cite{Klappenecker_2005}, we obtain the second case.
An explicit construction of a localization protocol is given in the paragraph just after Theorem~\ref{thm:lcalizable=nice}.

Now, we consider the case where $\{\dket[M_i]\}_{i=1,\cdots,4}$ is a product basis.
By straightforward calculation, we find that any two-qubit orthonormal product basis is LU-equivalent to either 
\begin{eqnarray}
    \{\ket[00],\ket[01],\ket[1e_0],\ket[1e_1]\}\\\
    \mathrm{or}\ \{\ket[00],\ket[10],\ket[e_01],\ket[e_11]\},
\end{eqnarray}
where $\{\ket[e_0],\ket[e_1]\}\subseteq\mathbb{C}^2$ is an orthonormal basis.
Since the set of localizable POVMs is symmetric under the permutation of two parties, we focus on the second case, i.e., 
\begin{equation}
    \label{eq:assumption_M}
    \{M_i\}_{i}=\{\ket[0]\bra[0],\ket[1]\bra[0],\ket[e_0]\bra[1],\ket[e_1]\bra[1]\},
\end{equation}
and prove that
$\{\dket[M_i]\}_{i}$ can be localized by $\dket[R]$ if and only if $\{\ketbra[e_0],\ketbra[e_1]\}=\{\ketbra[0],\ketbra[1]\}$ or $\{\ketbra[e_0],\ketbra[e_1]\}=R_z(\theta)\{\ketbra[+],\ketbra[-]\}R_z(\theta)^\dagger$ with $R_z(\theta)=\ketbra[0]+e^{i\theta}\ketbra[1]$, which completes the proof.

Lemma \ref{lem:rank-1_localizability} implies that $\{\op[M_i]\}_{i=1,\cdots,4}$ can be localized by $\dket[R]$ if and only if there exist rank-1 non-redundant POVMs $\{\op[A_a]\}_{a\in X}$ and $\{\op[B_b]\}_{b\in Y}$ such that for any $a\in X$ and $b\in Y$,
\begin{equation}
\label{eq:localization_cond}
    A_aR^*B_b\propto M_{f(a,b)},
\end{equation}
where $f:X\times Y\rightarrow Z$ and $Z=\{1,2,3,4\}$.

First, we consider the case where $\dket[R]$ is a product state. In this case, we can let $R^*=\ket[x]\bra[y]$ by using unit vectors $\ket[x],\ket[y]\in\mathbb{C}^2$. Since the completeness of the POVM $\{\op[A_a]\}_{a\in X}$ implies that $\{A_a\}_{a\in X}$ spans $\linop[\mathbb{C}^2]$, there exists $a\in X$ such that $A_a\ket[x]\neq0$. Since $\{B_b\}_{b\in Y}$ spans  $\linop[\mathbb{C}^2]$ due to the same argument, we find that 
\begin{equation}
    \vspan{\{M_{f(a,b)}\}_{b\in Y'}}=\{A_a\ket[x]\bra[z]:\ket[z]\in\mathbb{C}^2\},
\end{equation}
where $Y':=\{b\in Y:\bra[y]B_b\neq0\}$.
This holds if and only if $\{\ketbra[e_0],\ketbra[e_1]\}=\{\ketbra[0],\ketbra[1]\}$ in Eq.~\eqref{eq:assumption_M}.
In the following, we assume that $\dket[R]$ is an entangled state, i.e., $\mathrm{rank}(R^*)=2$.

Second, we consider the case where there exists $b\in Y$ such that $\mathrm{rank}(B_b)=2$.
Since $\{A_aR^*B_b\}_a\subseteq\cup_i(\mathbb{C}M_i)\Leftrightarrow\{A_a\}_a\subseteq\cup_i(\mathbb{C}M_iB_b^{-1}R^{*-1})$ and $\{\op[A_a]\}_{a\in X}$ is assumed to be a non-redundant POVM, we obtain $X=Z$ and
\begin{equation}
    \{A_aR^*B_b\}_{a\in X}=\{\alpha_iM_i\}_{i\in Z},
\end{equation}
where $\alpha_i\in\mathbb{C}^\times$. We relabel $a$ and identify $a\in X$ and $i\in Z$, i.e., $A_aR^*B_b=\alpha_aM_a$.
From Eq.~\eqref{eq:localization_cond}, we find that $\{\hat{B}_{b'}:=B_b^{-1}B_{b'}\}_{b'\in Y}$ satisfies that for any $a\in X$ and $b'\in Y$,
\begin{equation}
    M_a\hat{B}_{b'}=\alpha_a^{-1}A_aR^*B_bB_b^{-1}B_{b'}\propto A_aR^*B_{b'}\propto M_{f(a,b')}.
\end{equation}
Since $\{\hat{B}_{b'}\}_{b'\in Y}$ spans $\linop[\mathbb{C}^2]$ due to the completeness of the POVM $\{\op[B_b]\}_{b\in Y}$, we find that 
\begin{equation}
    \vspan{\{M_{f(a,b')}\}_{b'\in Y'}}=\{\ket[x]\bra[z]:\ket[x]\in\range{M_a},\ket[z]\in\mathbb{C}^2\},
\end{equation}
where $Y':=\{b'\in Y:M_a\hat{B}_b'\neq0\}$.
This holds if and only if $\{\ketbra[e_0],\ketbra[e_1]\}=\{\ketbra[0],\ketbra[1]\}$ in Eq.~\eqref{eq:assumption_M}.

Third, we consider the case where there exists $a\in X$ such that $\mathrm{rank}(A_a)=2$. Since $\{A_aR^*B_b\}_b\subseteq\cup_i(\mathbb{C}M_i)\Leftrightarrow\{B_b\}_b\subseteq\cup_i(\mathbb{C}R^{*-1}A_a^{-1}M_i)$ and $\{\op[B_b]\}_{b\in Y}$ is assumed to be a non-redundant POVM, we obtain $Y=Z$ and
\begin{equation}
    \{A_aR^*B_b\}_{b\in Y}=\{\alpha_iM_i\}_{i\in Z},
\end{equation}
where $\alpha_i\in\mathbb{C}^\times$. We relabel $b$ and identify $b\in Y$ and $i\in Z$, i.e., $A_aR^*B_b=\alpha_bM_b$.
From Eq.~\eqref{eq:localization_cond}, we find that $\{\hat{A}_{a'}:=A_{a'}A_a^{-1}\}_{a'\in X}$ satisfies that for any $a'\in X$ and $b\in Y$, 
\begin{equation}
    \hat{A}_{a'}M_b=A_{a'}A_a^{-1}\alpha_b^{-1}A_aR^*B_b\propto A_{a'}R^*B_b\propto M_{f(a',b)}.
\end{equation}
This implies that $\hat{A}_{a'}\ket[i]\in\mathbb{C}\ket[0]\cup\mathbb{C}\ket[1]$ and 
$\hat{A}_{a'}\ket[e_i]\in\mathbb{C}\ket[e_0]\cup\mathbb{C}\ket[e_1]$ for $i\in\{0,1\}$.
Thus, we can find that
\begin{eqnarray}
    \hat{A}_{a'}&\in&\bigcup_{\alpha,\beta\in\mathbb{C}}\left\{
    \begin{pmatrix}
        \alpha&\beta\\
        0&0
    \end{pmatrix},
    \begin{pmatrix}
        \alpha&0\\
        0&\beta
    \end{pmatrix},
    \begin{pmatrix}
        0&\beta\\
        \alpha&0
    \end{pmatrix},
    \begin{pmatrix}
        0&0\\
        \alpha&\beta
    \end{pmatrix}
    \right\},\nonumber\\
     \label{eq:compbasis}\\
    U\hat{A}_{a'}U^\dagger
    &\in&\bigcup_{\alpha,\beta\in\mathbb{C}}\left\{
    \begin{pmatrix}
        \alpha&\beta\\
        0&0
    \end{pmatrix},
    \begin{pmatrix}
        \alpha&0\\
        0&\beta
    \end{pmatrix},
    \begin{pmatrix}
        0&\beta\\
        \alpha&0
    \end{pmatrix},
    \begin{pmatrix}
        0&0\\
        \alpha&\beta
    \end{pmatrix}
    \right\}, \nonumber\\\label{eq:twistbasis} 
\end{eqnarray}
where we use the matrix representation of $\hat{A}_{a'}$ with respect to the computational basis and $U_{ij}=\braket{e_i}{j}$ is a unitary matrix.
If there exists $a'$ such that $\hat{A}_{a'}=
\begin{pmatrix}
        \alpha&\beta\\
        0&0
    \end{pmatrix}=
    \begin{pmatrix}
        1\\0
    \end{pmatrix}
    \begin{pmatrix}
        \alpha&\beta
    \end{pmatrix}(\neq 0)$ or $\hat{A}_{a'}=\begin{pmatrix}
        0&0\\
        \alpha&\beta
    \end{pmatrix}=
    \begin{pmatrix}
        0\\1
    \end{pmatrix}
    \begin{pmatrix}
        \alpha&\beta
    \end{pmatrix}(\neq 0)$, 
    $U\hat{A}_{a'}U^\dagger$ is either $\begin{pmatrix}
        \alpha&\beta\\
        0&0
    \end{pmatrix}$ or $\begin{pmatrix}
        0&0\\
        \alpha&\beta
    \end{pmatrix}$ since the rank of $\hat{A}_{a'}$ and $U\hat{A}_{a'}U^\dagger$ is $1$.
    This implies that $U\begin{pmatrix}
        1\\0
    \end{pmatrix}\in\mathbb{C}\begin{pmatrix}
        1\\0
    \end{pmatrix}\cup\mathbb{C}\begin{pmatrix}
        0\\1
    \end{pmatrix}$ and $U\begin{pmatrix}
        0\\1
    \end{pmatrix}\in\mathbb{C}\begin{pmatrix}
        1\\0
    \end{pmatrix}\cup\mathbb{C}\begin{pmatrix}
        0\\1
    \end{pmatrix}$. 
    Thus, we obtain that $U=\begin{pmatrix}
        \omega&0\\
        0&\omega'
    \end{pmatrix}$ or $U=\begin{pmatrix}
        0&\omega\\
        \omega'&0
    \end{pmatrix}$, where $|\omega|=|\omega'|=1$. 
    This implies that $\{\ketbra[e_0],\ketbra[e_1]\}=\{\ketbra[0],\ketbra[1]\}$.
    Otherwise, we can assume that
    \begin{eqnarray}
            \hat{A}_{a'}&\in&\bigcup_{\alpha,\beta\in\mathbb{C}^\times}\left\{
    \begin{pmatrix}
        \alpha&0\\
        0&\beta
    \end{pmatrix},
    \begin{pmatrix}
        0&\beta\\
        \alpha&0
    \end{pmatrix}
    \right\},\\ 
    \label{eq:complement_basis_cond}
    U\hat{A}_{a'}U^\dagger&\in&\bigcup_{\alpha,\beta\in\mathbb{C}^\times}\left\{
    \begin{pmatrix}
        \alpha&0\\
        0&\beta
    \end{pmatrix},
    \begin{pmatrix}
        0&\beta\\
        \alpha&0
    \end{pmatrix}
    \right\}
    \end{eqnarray}
    for all $a'\in X$, where we used the observation that $\mathrm{rank}(\hat{A}_{a'})=2$ and the unitary transformation does not change the matrix rank. Since $\vspan{\{\hat{A}_{a'}\}_{a'\in X}}=\linop[\mathbb{C}^2]$, there exists $a'\in X$ such that $\hat{A}_{a'}=a\begin{pmatrix}
        1&0\\
        0&1
    \end{pmatrix}+b\begin{pmatrix}
        1&0\\
        0&-1
    \end{pmatrix}$ with $b\in\mathbb{C}^\times$.
    If $a\neq0$, Eq.~\eqref{eq:complement_basis_cond} implies that
    $U\begin{pmatrix}
        1&0\\
        0&-1
    \end{pmatrix}U^\dagger=\begin{pmatrix}
        \alpha&0\\
        0&\beta
    \end{pmatrix}$ since $U\begin{pmatrix}
        1&0\\
        0&1
    \end{pmatrix}U^\dagger=\begin{pmatrix}
        1&0\\
        0&1
    \end{pmatrix}$.
    If $a=0$, Eq.~\eqref{eq:complement_basis_cond} implies that $U\begin{pmatrix}
        1&0\\
        0&-1
    \end{pmatrix}U^\dagger\in\left\{
    \begin{pmatrix}
        \alpha&0\\
        0&\beta
    \end{pmatrix},
    \begin{pmatrix}
        0&\beta\\
        \alpha&0
    \end{pmatrix}
    \right\}$.
    These imply that
    \begin{equation}
        U\begin{pmatrix}
        1&0\\
        0&-1
    \end{pmatrix}U^\dagger\in\bigcup_{\omega\in\mathbb{C},|\omega|=1}\left\{\begin{pmatrix}
        1&0\\
        0&-1
    \end{pmatrix},
    \begin{pmatrix}
        -1&0\\
        0&1
    \end{pmatrix},\begin{pmatrix}
        0&\omega\\
        \omega^*&0
    \end{pmatrix}\right\}
    \end{equation}
    since a unitary transformation does not change eigenvalues.
    Thus, we obtain that $U\in\left\{
    \begin{pmatrix}
        \omega&0\\
        0&\omega'
    \end{pmatrix},
    \begin{pmatrix}
        0&\omega\\
        \omega'&0
    \end{pmatrix},
    \frac{\omega''}{\sqrt{2}}\begin{pmatrix}
        \omega'&-\omega^*\\
        \omega&\omega'^*
    \end{pmatrix}\right\}$, where $|\omega|=|\omega'|=|\omega''|=1$. Thus, we obtain $\{\ketbra[e_0],\ketbra[e_1]\}=\{\ketbra[0],\ketbra[1]\}$ or $\{\ketbra[e_0],\ketbra[e_1]\}=R_z(\theta)\{\ketbra[+],\ketbra[-]\}R_z(\theta)^\dagger$.

    Fourth, we consider that case where $\mathrm{rank}(A_a)=\mathrm{rank}(B_b)=1$ for all $a\in X$ and $b\in Y$.
    Since the completeness of the POVM $\{\op[A_a]\}_{a\in X}$ implies that $\{A_a\}_{a\in X}$ spans $\linop[\mathbb{C}^2]$, there exists $a\in X$ such that $A_aR^*\neq0$. Since $\{B_b\}_{b\in Y}$ spans  $\linop[\mathbb{C}^2]$ due to the same argument, we find that 
    \begin{equation}
        \vspan{\{M_{f(a,b)}\}_{b\in Y'}}=\{\ket[x]\bra[z]:\ket[x]\in\range{A_a},\ket[z]\in\mathbb{C}^2\},
    \end{equation}
    where $Y':=\{b\in Y:A_aR^*B_b\neq0\}$.
    This holds if and only if $\{\ketbra[e_0],\ketbra[e_1]\}=\{\ketbra[0],\ketbra[1]\}$ in Eq.~\eqref{eq:assumption_M}.

$\qed$

\end{document}